%% file: CPH_AM.tex
\documentclass{article}

\usepackage{arxiv}

\usepackage[utf8]{inputenc} 
\usepackage[T1]{fontenc}    
\usepackage{lmodern}        
\usepackage{hyperref}       
\usepackage{url}            
\usepackage{booktabs}       
\usepackage{amsfonts}       
\usepackage{nicefrac}       
\usepackage{microtype}      
\usepackage{graphicx}

\title{A scalable and flexible Cox proportional hazards model for
high-dimensional survival prediction and functional selection}

\author{
    Boyi Guo
    \thanks{Corresponding authors
\textbackslash{}\href{mailto:email\%7Bboyiguo1@uab.edu}{\nolinkurl{email\{boyiguo1@uab.edu}}\}}
   \\
    Department of Biostatistics \\
    University of Alabama at Birmingham \\
  Birmingham, AL \\
  \texttt{} \\
   \And
    Nengjun Yi
    \thanks{Corresponding authors
\textbackslash{}\href{mailto:email\%7Bnyi@uab.edu}{\nolinkurl{email\{nyi@uab.edu}}\}}
   \\
    Department of Biostatistics \\
    University of Alabama at Birmingham \\
  Birmingham, AL \\
  \texttt{} \\
  }

\usepackage{amsmath}
\begin{document}
\maketitle

\begin{abstract}
Cox proportional hazards model is one of the most popular models in
biomedical data analysis. There have been continuing efforts to improve
the flexibility of such models for complex signal detection, for
example, via additive functions. Nevertheless, the task to extend Cox
additive models to accomodate high-dimensional data is nontrivial. When
estimating additive functions, commonly used group sparse regularization
may introduce excess smoothing shrinkage on additive functions, damaging
predictive performance. Moreover, an ``all-in-all-out'' approach makes
functional selection challenging to answer if nonlinear effects exist.
We develop an additive Cox PH model to address these challenges in
high-dimensional data analysis. Notably, we impose a novel
spike-and-slab LASSO prior that motivates the bi-level functional
selection on additive functions. A scalable and deterministic algorithm,
EM-Coordinate Descent, is designed for scalable model fitting. We
compare the predictive and computational performance against
state-of-the-art models in simulation studies and metabolomics data
analysis. The proposed model is broadly applicable to various fields of
research, e.g.~genomics and population health, via freely available R
package BHAM (\url{https://boyiguo1.github.io/BHAM/})
\end{abstract}

\keywords{
    Cox Model; Spike-and-Slab; Scalable; Machine Learning; Additive
    Models;
  }

\newcommand{\pr}{\text{Pr}}
\newcommand{\bs}[1]{\boldsymbol{#1}}
\newcommand{\tp}{*}
\newcommand{\simiid}{\overset{\text{iid}}{\sim}}

\section{Introduction}

Cox proportional hazards model (Cox model hereafter) is one of the most
popular analytic choices to analyze right-censored time-to-event data.
Cox model describes the change of risk, modeled as hazard, on the
multiplicative scale and provides easily interpretable insights in
disease etiology. When a variable enters the Cox model as a predictors,
an implicit assumption is imposed: the effect of the predictor are
linear. The linear assumption is restrictive and doubtful in many case.
A viable solution is to replace each variable with its functional form
such that the modified Cox model retains its interpretability while
gaining flexibility to model complex signals. The Cox model with
additive components (hereafter referred to as Cox additive model or CAM
for short) has many successful applications in biomedical research, for
example, dose-response curve modeling \cite{Steenland2004}, disease
progrognostic \cite{gray1992} to name a few. To clarify, the Cox
additive model differs from the additive hazards model \cite{aalen1980}
even both models leverage additive functions of predictors. The two
models answer different scientific questions: the Cox additive model
measures hazard change on the multiplicative scale and produces risk
ratio interpretation; the additive hazards model measures hazard change
on the additive scale and produces risk difference interpretation. In
this manuscript, we limit our focus to the Cox additive model, and defer
interested readers to \cite{lin1997} for a review of the additive
hazards model.

Splines function are one of the most popular choices for additive
functions in CAM, because they are mathematically simple, smooth across
the range, and hence provide easy interpretation. Mathematically, a
spline funciton is a piece-wise polynomial function with continuity
conditions imposed on the function. \cite{Wood2017} To apply spline
functions in CAM, one can easily replace each predictor with the matrix
form of its corresponding spline function which is known. The estimation
of coefficients follows the same procedure as fitting an ordinary Cox
model. This approach is called regression spline. Nevertheless, if
spline functions are overparameterized, i.e.~using more than necessary
degree of smoothness, regression spline models incline to be overfitted
and the estimated functions are very wiggly. To make data-driven
decision on the degrees of smoothness, smoothing penalty are applied,
resulting the smoothing spline model \cite{reinsch1967}.

Large volumes of biomedical data motivate the development of
high-dimensional statistics. Here, we define high-dimensional statistics
as the analytic models to address analyses where the number of
predictors/dimensions is close, if not more than, the sample size
(\(p>>n\)), commonly seen in -omics data and high-resolution image data.
There have been sustaining efforts to extend Cox model to accommodate
high-dimensional setting
\cite{tibshirani1997, fan2002, zhang2007, wu2012, bradic2011, fan2010},
with a few allowing a small subset of predictors to be modeled
nonparametrically\cite{Du2010} (also known semiparametric regression).
The semiparametric regression approaches improve the flexibility of
high-dimensional Cox models, but the improvement can be limited when the
knowledge about predictors' linearity is lacking, for example during the
exploratory data analysis of genomics studies. A nonparametric
regression approach is highly sought-after for full flexibility and
autonomy.

A technical gap exists when attempting to extend the semiparametric
regression to the nonparametric regression under the Cox proportional
hazard framework for high-dimensional data analysis. Semiparametric
regressions assume all additive functions included are necessary and
only address the smoothing of these functions. This strategy would fail
when applied to nonparametric regressions, as the selection of additive
functions are necessary. In other words, under the nonparametric
setting, one needs to address the sparsity of additive functions in
addition to the smoothness of effective functions. One school of
thoughts is to employ group sparse penalty on the coefficients of
additive functions: \cite{leng2006} proposed a functional analogue of
LASSO penalty;\cite{Lian2013} proposed to uses SCAD penalties to
selecting non-parametric component functions;\cite{Yang2018} proposed a
feature screening procedure for the additive cox model. These proposals
can serve as effective variable selection tools, but are susceptible to
inaccurate risk prediction. When constructing model penalty, function
smoothness are not considered. Such strategies are not recommended in
the high-dimensional generalized additive model literature, as models
are prone to wiggle estimations when underlying signal is smooth.
\cite{Meier2009} In addition, these globally penalized models ignore the
fact that each additive function can have different degrees of
smoothness, and hence introduce an over-smoothed solution.
\cite{scheipl2013} There are also other approaches provides flexible
solutions to model survival outcome in the high-dimensional setting.
\cite{bender2018} proposed to model survival outcome with piece-wise
exponential models, which is based on the idea of fit a Poisson GLM with
transformed survival outcomes. Nevertheless, this method is not
computationally efficient and vulnerable to convergence problem due to
the numeric calculations. \cite{wu2019} extended the trend filtering
model to the survival setting. The solution is a series of step
functions, which would not be optimal if underlying functions are
assumed to be smooth.

In this article, we introduce the two-part spike-and-slab LASSO (SSL)
prior for smooth functions \cite{guo2022} to the high-dimensional Cox
additive model literature. Specially, the two-part SSL prior
simultaneously addresses signal sparsity and function smoothness by
allowing separate adaptive shrinkages on linear and nonlinear components
of additive functions. Built upon the premise that better signal
estimation produce better outcome prediction, we adopt the smoothing
penalty concept via design matrix reparameterization to encourage
accurate function estimation. In addition, the two-part SSL prior
motivates the bi-level functional selection. Here, we define the
bi-level selection as the decision-making on if a variable should be
included in the model and if the variable has linear versus nonlinear
effect. To the best of our knowledge, we are the first to apply
spike-and-slab LASSO prior in the context of high-dimensional Cox
additive model for the purposes of simultaneous survival prediction and
functional selection. Meanwhile, fitting a Bayesian hierarchical model
for high-dimensional data can be computationally intimidating. Hence, we
develop a scalable EM-Coordinate Descent algorithm and provide
user-friendly implementation in the open-source software environment
R\cite{R2021}. The implementation is freely available via
\url{https://github.com/boyiguo1/BHAM}. The proposed framework
contributes a flexible and efficient solution to high-dimensional
molecular and clinical data analysis.

\section{Cox Proportional Hazards Additive Model}

For each individual indexed with \(i\), we collect the the covariate
variables
\(\boldsymbol{X}_i = (X_{i1}, \dots, X_{ij}) \in \mathbb{R}^{p}\) and
the survival outcome \((T_i, D_i) \in \{\mathbb{R}^+ \times \{0,1\}\}\)
where \(T_i\) is the observed survival time, and \(D_i\) is a binary
censoring indicator. \(D_i\) takes the value 1 when the event of
interest happens at the survival time \(T_i\), and takes the value 0
when censoring happens. We assume non-informative right censoring, no
competing risk or multiple occurrence of the event. The Cox proportional
hazards model with additive functions
\(B_j(x_j) = \boldsymbol{\beta}_j \boldsymbol{X}_j\) is formulated as \[
h(t_i|\boldsymbol{x}_i) = h_0(t_i)exp(\sum\limits^p_{j=1}B_j(x_{ij})) = h_0(t_i) \exp(\sum\limits^p_{j=1}\boldsymbol{\beta}^T_j \boldsymbol{X}_{ij}).
\] For the purpose of identifiability, we impose an constraint on each
additive function such that \(E[B_j(x_j)] = 0\). \(h(t_i)\) is the
hazard function and \(h_0(t_i)\) is the baseline hazard function. The
hazard function \(h(t)\) describes the instantaneous rate of event
occurrence among people who are still at risk at the moment. To note,
different from GLMs, there is no intercept term necessary as the
baseline hazard function \(h_0(t)\) estimates an reference level of
survival risk. We defer to \cite{Klein2003, Ibrahim2001} for full
treatment of Cox proportional hazards models.

We consider maximizing the partial log-likelihood function
\cite{cox1972} for Cox model fitting, mathematically,
\begin{equation}\label{eq:partial-likelihood}
pl(\boldsymbol{\beta}) = \sum\limits^n_{i=1}d_i\log\frac{\exp{\boldsymbol{x}_i \boldsymbol{\beta}}}{\sum_{j\in R(t_i)} \exp(\boldsymbol{x}_j \boldsymbol{\beta})},
\end{equation} where \(R(t_i)\) denotes the risk set at time \(t_i\),
i.e.~the set of all patients who still survived prior to time \(t_i\).
When tied failure or censoring time exists, a modified partial
log-likelihood function can be used \cite{Efron1977}. Conditional on
\(\boldsymbol{\beta}\), the baseline hazard function \(h_0(t_i)\) can be
estimated using Breslow estimator \cite{breslow1974}, \[
\hat h_0(t_i|\beta) = d_i/\sum\limits_{i^\prime \in R(t_i)} exp(X_{i^\prime}\beta).
\]

To encourage proper smoothing of the functions, we adopt the idea of
smoothing penalties from smoothing spline models \cite{Wood2017}.
Smoothing penalty facilitate the data-driven estimation of function
smoothness and is mathematically defined as the integrated squared
second derivative of smooth functions. While it is hard to directly
integrate smoothing penalty with sparsity penalty in high-dimensional
settings, \cite{Marra2011} proposed a reparameterization to absorb
smoothing penalty to the design matrix. Given the known smoothing
penalty matrix \(\boldsymbol{S}_j\) is symmetric and positive
semi-definite for univariate additive functions \(B_j(x_j)\), we apply
eigendecomposition on the penalty matrix
\(\boldsymbol{S} = \boldsymbol{U} \boldsymbol{D} \boldsymbol{U}^T\),
where eigenvectors and eigenvalues are arranged in the matrices
\(\boldsymbol{U}\) and \(\boldsymbol{D}\) respectively. The zero
eigenvalues and the corresponding eigenvectors span the linear space of
the smoothing function, which allows us to separate the linear space
from the smoothing function. By multiplying the design matrix \(X\) and
eigenvector matrix \(U\) and properly scaling by the eigenvalues, we can
have a new design matrix such that the smoothing function of variable
\(x_j\) can be written as \[
B_j(x_j) = B_j^0(x_j) + B_j^*(x_j) = \beta_j X^0_j + \boldsymbol{\beta_j^*}^T \boldsymbol{X}_j^*,
\] where
\(\boldsymbol{X}_j \equiv \begin{bmatrix} X_j^0 : \boldsymbol{X}_j^{*} \end{bmatrix}\)
as the basis function matrix for the \(j\)th variable; the coefficients
\(\boldsymbol{\beta}_j \equiv \begin{bmatrix} \beta_j : \boldsymbol{\beta}^*_j \end{bmatrix}\)
is an augmentation of the coefficient scalar \(\beta_j\) of linear space
and the coefficient vector \(\boldsymbol{\beta}^*_j\) of non-linear
space. The dimension for \(\boldsymbol{X}_j^{*}\) and
\(\boldsymbol{\beta}^*_j\) is \(K_j\). The reparameterization allows the
coefficients to be on the same scale and encourages separate
consideration of the linear and nonlinear spaces of an additive
function, which motivates the proposed two-part prior.

\subsection{Two-part Spike-and-slab LASSO Prior for Smooth Functions}

To model each additive function, we propose the two-part spike-and-slab
LASSO prior under the additive Cox proportional hazard framework. The
proposed prior is an extension of the previous spike-and-slab Lasso
prior for group predictors \cite{tang2019}, and has been applied to the
GAM settings \cite{guo2022}. The proposed prior provides three-folded
advantages: 1) data-driven estimation of function smoothness via
adaptive shrinkage; 2) natural bi-level selection of additive functions
without hypothesis testing or thresholding; 3) efficient model fitting
with a scalable algorithm.

To recall, the spike-and-slab LASSO prior
\cite{rockova2018a, rockova2018b} is a mixture double exponential prior
with a spike density \(DE(0, s_0)\) for small effects and a slab density
\(DE(0, s_1)\) for large effect (\(0 < s_0 < s_1\)). Like any other
spike and slab priors, the spike is to contain the minimum to zero
effects, while the slab is to allow large effects. The scale parameters
\(s_0\) and \(s_1\) are considered given and can be optimized via
cross-validation. A latent indicator variable \(\gamma_j \in \{0,1\}\)
controls if the predictor is include in the model or not.
Mathematically, the spike-and-slab LASSO prior is expressed as \[
\beta|\gamma \sim (1-\gamma) DE(0, s_0) + \gamma DE(0, s_1), 0 < s_0 < s_1.
\] To accommodate the group structure of the predictors,
\cite{tang2018, tang2019} proposed the group spike-and-slab LASSO prior
by imposing a group specific Bernoulli distribution on the indicator
variables, \[
\gamma_j|\theta_g \sim Bin(1, \theta_g).
\] The probability parameter \(\theta_g\) of group \(g\) allows
information borrowing across different predictors in the same group. The
group-specific hyper prior is built on the premise that if one predictor
in the group is included in the model, the rest of the predictors are
more likely to be in the model. The spike-and-slab LASSO prior can be
seen as a special case of the group spike-and-slab LASSO prior where the
size of each group is one.

In the proposed two-part spike-and-slab LASSO prior for smooth function,
we leverage the previous group spike-and-slab LASSO prior and make
modification, such as group latent indicator and effect hierarchy
principle. Specifically, we impose conditionally independent group SSLs
on the linear and nonlinear components of a smoothing function. This
accounts the natural group structure among additive function bases.
Given the model matrix of the predictor \(\boldsymbol{X}_j\) after the
reparameterization step, we have the linear and nonlinear component of
the smooth function \(X_j^0\) and \({\boldsymbol{X}_j}^*\). The
corresponding coefficients are
\(\boldsymbol{\beta}_j = [\beta_j:\boldsymbol{\beta}_j^*]\). We impose
the group SSL priors on the linear and nonlinear components
respectively, \begin{align}
  \beta_{j} | \gamma_{j},s_0,s_1 &\sim (1-\gamma_{j}) DE(0, s_0) + \gamma_{j} DE(0, s_1)\nonumber \\
  \beta^*_{jk} | \gamma^*_{j},s_0,s_1 &\overset{\text{iid}}{\sim}(1-\gamma_{j}^*) DE(0, s_0) + \gamma_{j}^*DE(0, s_1), k=1,\dots, K_j.
\end{align} To note, we assume that the linear components are
one-dimensional and hence using the special case, SSL prior, for
simplicity. We make slight modification on the previous group SSL: we
have all coefficients of the nonlinear components, i.e.~within the same
group, to share the same group latent binary indicator. This encourages
the inclusion of nonlinear components all together and hence, the
bi-level selection. Meanwhile, we see that the all the nonlinear
components should have the same magnitude of shrinkage after
reparameterization, particularly the scaling. The group spike-and-slab
Prior for the nonlinear component can be considered as the smoothing
penalty in the smoothing spline setting.

While the two latent indicators \(\gamma_j\) and \(\gamma_j^*\) controls
the inclusion of the linear and nonlinear components of a smooth
function, we still need to set up some ordering of the inclusion. For
example, it is often to assume that the lower-order effects are more
likely to be active than the high-reorder effects (referred to as
\textit{effect hierachy} \cite{chipman2006}). In order to implement the
effect hierarchy principle in the bi-level selection, we further impose
a dependent structure on the latent indicators, \begin{align}
&\gamma_{j} | \theta_j \sim Bin(1, \theta_j) & & 
&\gamma_{j}^*| \gamma_{j}, \theta_j \sim Bin(1, \gamma_{j}\theta_j).
\end{align} This is the inclusion of the nonlinear component depends on
the inclusion of the linear component. To note, one can easily relax the
effect hierarchy by having the two latent indicators be independent
condition on the inclusion probability parameter \(\theta_j\). The two
versions of the indicator prior could introduce trade-offs in the
variable selection (previously seen in \cite{guo2022}) and will be
discussed more in the Section 5. It is also possible to have the linear
and nonlinear coefficients shares the same indicators,
i.e.~\(\beta_{jk}^*\) also depends on \(\gamma_j\). However, this
approach would disable the bi-level selection ability and force the
nonlinear components uses sparsity shrinkage instead of smoothing
shrinkage. Hence, it would reduce to a more strict version of group
spike-and-slab LASSO where all coefficients in a group employs the same
shrinkage instead of locally adaptive shrinkage for linear effect and
nonlinear effects respectively, and hence it is not recommend here.

The rest of the proposed prior follows the spike-and-slab LASSO prior.
The inclusion probability parameter \(\theta_j\) independently and
identically follows a \(Beta(a, b)\) distribution. One can consider a
special case of the \(Beta\) distribution, \(Uniform(0,1)\), for
simplicity. The prior distribution of the inclusion probability
parameter motivates the self-adapative shrinkage for signal sparsity and
functional smoothness based on the data. In addition, being a conjugate
prior of the binomial distribution, the \(Beta\) prior can provide a
closed-form solution in the following model fitting algorithm and
mitigate some computational burdens. To note, when the variable have
large effects in any of the bases, the parameter \(\theta_j\) will be
estimated large, which in turn encourages the model to include the rest
of bases.

\subsection{EM-Coordinate Descent Algorithm for Scalable Model Fitting}

Parsimonious computation is always encouraged in high-dimension data
analysis. Many Bayesian methods lose their advantages over penalized
models because of their reliance on the computationally prohibitive
model fitting algorithms. Previous Bayesian additive models relies
heavily on Markov chain Monte Carlo algorithms to approximate posterior
distribution of parameters. An economic alternative is to use
optimization-based algorithms, e.g.~EM procedure, to derive the maximum
a posteriori estimates, previously seen in \cite{bai2021, guo2022} for
Bayesian generalized additive models. Nevertheless, these algorithms do
not apply to the Cox model.

We develop a fast deterministic algorithm to fit the proposed
spike-and-slab LASSO Cox additive model. The algorithm is an extension
of the previously proposed EM-coordinate descent algorithm for group
spike-and-slab LASSO Cox \cite{tang2019}. Specifically, we first
formulate the spike-and-slab LASSO prior as a double exponential prior
with a conditional scale parameter. Next, we leverage the relationship
between posterior density function and penalized likelihood function,
\(l_1\) penalized specifically, and maximize the posterior density
function via coordinate descent algorithm. The feasibility of the
optimization process conditions on some nuisance parameters, and hence,
we use the Expectation-Maximization procedure to iteratively update the
parameters of interests until convergence. We see that similar
strategies achieve great computational convenience compared to versions
of Monte Carlo Markov Chain algorithms in the high-dimensional data
analysis literature, for example \cite{rockova2014, tang2017, guo2022}
to name a few.

In the proposed algorithm, our objective is to find the parameters of
interests \(\Theta = \{\boldsymbol{\beta}, \boldsymbol{\theta}\}\) that
maximize the log joint posterior density of
\(\Theta, \boldsymbol{\gamma}\). In Bayesian survival analysis, it is
common to approximate posterior density with the product of partial
likelihood function in Equation \eqref{eq:partial-likelihood} and
marginal priors. \cite{sinha2003} Hence, our objective function (up to
additive constants) is expressed mathematically,
\begin{align}\label{eq:bcam_obj_fun}
\max\limits_{\Theta, \boldsymbol{\gamma}} \log f(\boldsymbol{\beta}, \boldsymbol{\theta}, \boldsymbol{\gamma }|t, d) &= \log pl(\boldsymbol{\beta})+ \sum\limits_{j=1}^p\left[\log f(\beta_j|\gamma_j)+\sum\limits_{k=1}^{K_j} \log f(\beta^{*}_{jk}|\gamma^{*}_{j})\right]\nonumber\\
+&\sum\limits_{j=1}^{p} \left[ (\gamma_j+\gamma_{j}^{*})\log \theta_j + (2-\gamma_j-\gamma_{j}^{*}) \log (1-\theta_j)\right] +  \sum\limits_{j=1}^{p}\log f(\theta_j).
\end{align}

Given the latent inclusion indicator is binary and takes only 0 and 1 as
its value, a spike-and-slab LASSO prior, as well as the group version,
can be expressed as a double exponential prior whose scale parameter is
\(s_0\) when \(\gamma = 0\) and \(s_1\) when \(\gamma = 1\),
\begin{equation}
\beta|\gamma, s_0,s_1 \sim DE(0, (1-\gamma_{j}) s_0 + \gamma_{j} s_1). \nonumber
\end{equation} Leveraging the relationship between double exponential
prior and LASSO, the product of partial likelihood function and the
prior of \(\boldsymbol{\beta}\) can be viewed as an \(l_1\) penalized
partial likelihood function with penalty
\(\lambda_j = \{(1-\gamma_{j}) s_0 + \gamma_{j} s_1\}^{-1}\) and
\(\lambda_j^*= \{(1-\gamma_{j}^*) s_0 + \gamma_{j}^*s_1\}^{-1}\) for
\(\beta_j\) and \(\beta_{jk}^*\) and optimized with the coordinate
descent algorithm \cite{simon2011}. Nevertheless, the optimization
requires the knowledge of \(\boldsymbol{\gamma}\), which is unknown. To
address the problem, we treat the latent indicators
\(\boldsymbol{\gamma}\) as the ``missing data'' and use EM algorithm to
iteratively derive the maximum a postiori estimates. Notably, we
establish the expected log joint posterior density of
\(\Theta, \boldsymbol{\gamma}\) with respect to the latent indicators
conditioning on the parameters of interest estimated from previous
iteration
\(\Theta^{(t-1)}\).\footnote{The superscription $^{(t)}$ denotes the the parameter estimates at the $t$th iteration.}
Hence, we can calculate the equivalent \(l_1\) penalty at the \(t\)-th
iteration in the EM algorithm as
\(\lambda_j^{(t)} = \frac{1-p_{j}^{(t)}}{s_0} + \frac{p_{j}^{(t)}}{s_1}\)
and
\({\lambda_j^*}^{(t)} = \frac{1-{p_{j}^*}^{(t)}}{s_0} + \frac{{p_{j}^*}^{(t)}}{s_1}\)
for \(\beta_j\) and \(\beta_{jk}\), where
\(p_{j}^{(t)} \equiv \text{Pr}(\gamma_{j}=1|\Theta^{(t-1)})\) and
\(p_{j}^*\equiv \text{Pr}(\gamma^*_{j}=1|\Theta^{(t-1)})\). To note, for
computational convenience, we analytic integrate \(\gamma_j\) out of the
prior density of \(\gamma_j^*\). The two quantity \(p_j^{(t)}\) and
\({p_{j}^*}^{(t)}\) can be easily derived with Bayes' theorem and we
defer the derivation to \cite{guo2022}. With the expectation set up, we
can update \(\boldsymbol{\beta}^{(t)}\) with the coordinate descent
algorithm as previously described. Meanwhile, conditioning on
\(\boldsymbol{\gamma}\), the rest of components in Equation
\eqref{eq:bcam_obj_fun} can be optimized independently from the
penalized partial likelihood function. It is easy to update
\(\theta_j^{(t)}\) with a closed-form equation due to the conjugate
relationship, \begin{equation}\label{eq:update_theta}
\theta_j^{(t)} = \frac{p_j^{(t)} + {p_{j}^*}^{(t)} + a - 1 }{a + b}.
\end{equation} The E- and M- steps iterates until meeting the
convergence criterion \(|d^{(t)}-d^{(t-1)}|/(0.1+|d^{(t)}|)<\epsilon\),
where \(d^{(t)} = -2pl(\boldsymbol{\beta}^{(t)})\) and \(\epsilon\) is a
small value (say \(10^{-5}\))

Totally, the proposed EM-CD algorithm is summarized as follows:

\begin{enumerate}
\def\labelenumi{\arabic{enumi})}
\item
  Choose a starting value \(\boldsymbol{\beta}^{(0)}\) and
  \(\boldsymbol{\theta}^{(0)}\) for \(\boldsymbol{\beta}\) and
  \(\boldsymbol{\theta}\). For example, we can initialize
  \(\boldsymbol{\beta}^{(0)} = \boldsymbol{0}\) and
  \(\boldsymbol{\theta}^{(0)} = \boldsymbol{0}.5\)
\item
  Iterate over the E-step and M-step until convergence

  E-step: calculate \(E(\gamma_{j})\), \(E(\gamma^*_{j})\) and
  \(E({S}^{-1}_{j})\), \(E({S^*}^{-1}_{j})\) with estimates of
  \(\Theta^{(t-1)}\) from previous iteration

  M-step:

  \begin{enumerate}
  \def\labelenumii{\alph{enumii})}
  \item
    Update \(\boldsymbol{\beta}^{(t)}\) by optimizing the penalized
    likelihood function in Equation (xx) using the coordinate descent
    algorithm
  \item
    Update \(\boldsymbol{\theta}^{(t)}\) using the closed-form
    calculation in Equation \eqref{eq:update_theta}
  \end{enumerate}
\end{enumerate}

\subsubsection{Selecting Optimal Scale Values}

In the proposed two-part spike-and-slab LASSO prior, we assume the scale
parameters \(s_0, s_1\) known. As the model performance depends on the
values of the two scale parameters, we use cross-validation with respect
to a criteria of preference, for example the the partial log-likelihood,
concordance index, the survival curves, or the survival prediction
error, to decide their optimal values. Meanwhile, previous research
showed the value of the slab scale \(s_1\) has less impact on the final
model and is recommended to be set as a generally large value,
e.g.~\(s_1 = 0.5\), that provides no or weak shrinkage. \cite{tang2019}
Hence, instead of constructing a two-dimensional grid, we focus on
examining different values of the spike scale \(s_0\). Similar to the
LASSO implementation in the widely used R package \texttt{glmnet}, we
examine a sequence of models with different \(s_0\) values and have
users to choose from.

\section{Simulation Studies}

In this section, we evaluate the prediction performance of the proposed
model against three state-of-the-art Cox additive models, including mgcv
\cite{Wood2017}, component selection and smoothing operator (COSSO)
\cite{leng2006}, and adaptive COSSO \cite{storlie2011}. mgcv is the
implementation of generalized additive models with automatic smoothing,
and is the one of the most popular method to model nonlinear signals
under the Cox proportional hazard framework. To note, mgcv doesn't
support analyses when the number of parameters is larger than the sample
size, and would not work in the \(p>n\) scenario. COSSO and adaptive
COSSO is designed to solve the nonlinear effect modelling in the
high-dimensional setting. COSSO is one of the earliest additive model
that leverage the sparsity-smoothness penalty, and adaptive COSSO
improves COSSO by using adaptive weight for penalties aiming to relax
from the uniform shrinkage applied to all additive functions. The three
models of comparison are implemented with R packages \texttt{cosso}
2.1-1 \cite{R_cosso}, and \texttt{mgcv} 1.8-31 \cite{R_mgcv}
respectively. To make the evaluation fair, we control multiple
implementation factors that could alter the performance, including the
smoothing function and tuning of the models. We control the
dimensionality of the smoothing functions to 10 bases. We use the most
popular cubic spline as the choice of smoothing function for mgcv and
the proposed model. COSSO models do not provide any flexibility to
define smooth functions, and hence use the default choice. We use 5-fold
cross-validation to select the tuning parameter among 20 default
candidates except \texttt{mgcv} which uses generalized cross-validation
to select optimal model. The simulation study is conducted with R 4.1.0
on a high-performance 64-bit Linux platform with 48 cores of 2.70GHz
eight-core Intel Xeon E5-2680 processors and 24G of RAM per core.

\subsection{Data Generating Process}

To established a comprehensive understanding of the methods performance,
we consider multiple factors that are pivotal to high-dimensional data
analysis, nonlinear modeling and survival outcomes. We examine different
settings of signal sparsity (defined as the ratio of active variables
and total number of covariates), sample size, correlation structure of
the predictors, functional form of the underlying signals, and censoring
rate.

To describe the data generating process, we generate a total of 1200
data points, where 200 serves as the training data and 1000 serves as
the testing data. We consider the number of predictors \(p\) to be \{4,
10, 50, 100, 200\} while limiting the number of active predictors to be
4. We simulate the predictors \(\boldsymbol{X}\) from a multivariate
normal distribution MVN\(_{200\times p}(0, \Sigma)\). The variance
covariance matrix \(\Sigma\) follows a auto-regressive (AR) structure
with two possible order parameters, \{0, 0.5\}, where \(AR(0)\)
indicates the predictors are mutually independent. Among the all the
predictors, we choose the first four to be the active predictors,
i.e.~\(B_1(x) = (x+1)^2/5, B_2(x) = \exp(x+1)/25, B_3(x_3) = 3*sin(x)/2\),
and \(B_4(x) = (1.4*x+0.5)/2\). The rest of the predictors are inactive,
i.e.~\(B_j(x) = 0\) for \(j = 5, \dots, p.\) To simulate the the
survival response, we first generate the ``true'' survival time \(T_i\)
for each individual from a Weibull distribution with the scale parameter
1 and shape parameter 1.2 with the help of R package \texttt{simsurv}
1.0.0 \cite{R_simsurv}. We then generate the independent censoring time
\(C_i\) following a Weibull distribution with shape parameter 0.8. We
use \cite{wan2016} to estimate the scale parameter so that the censoring
rate is controlled at \{0.15, 0.3, 0.45\}. To note, numeric problems can
happen when estimating the scale parameter. Instead, we use the median
of other estimated scale parameters of the same simulation setting. The
observed censored survival time is the minimum of the ``true'' survival
and censoring time \(t_i= min(T_i, C_i)\). The censoring indicator was
set to be \(I(C_i>T_i)\).

In each iteration of the simulation process, we independently generate
the training and testing datasets following the previously described
data generation process. We use the training dataset to construct each
model of comparison and find the optimal model using 5-fold
cross-validation. Then we use the fitted model to make predictions for
the testing dataset and calculate the evaluation metrics, including
deviance and C-index.

\subsection{Simulation Results}

Across all the simulation settings, the empirical censoring rate is
controlled at the desired level (see Table \ref{tab:sim_cnr_prop}). As
previously mentioned, mgcv doesn't fit model when the number of
parameters exceeds the sample size, and hence ignored in p = \{100,200\}
evaluations. We also experience some programming errors when fitting
COSSO and adaptive COSSO models when p is small. The proposed method is
robust to programming errors in all examined settings. Our following
performance evaluation only use successful iterations. In the following
evaluation, we only summarize the performance from success runs.

\begin{figure}[h] 
\includegraphics{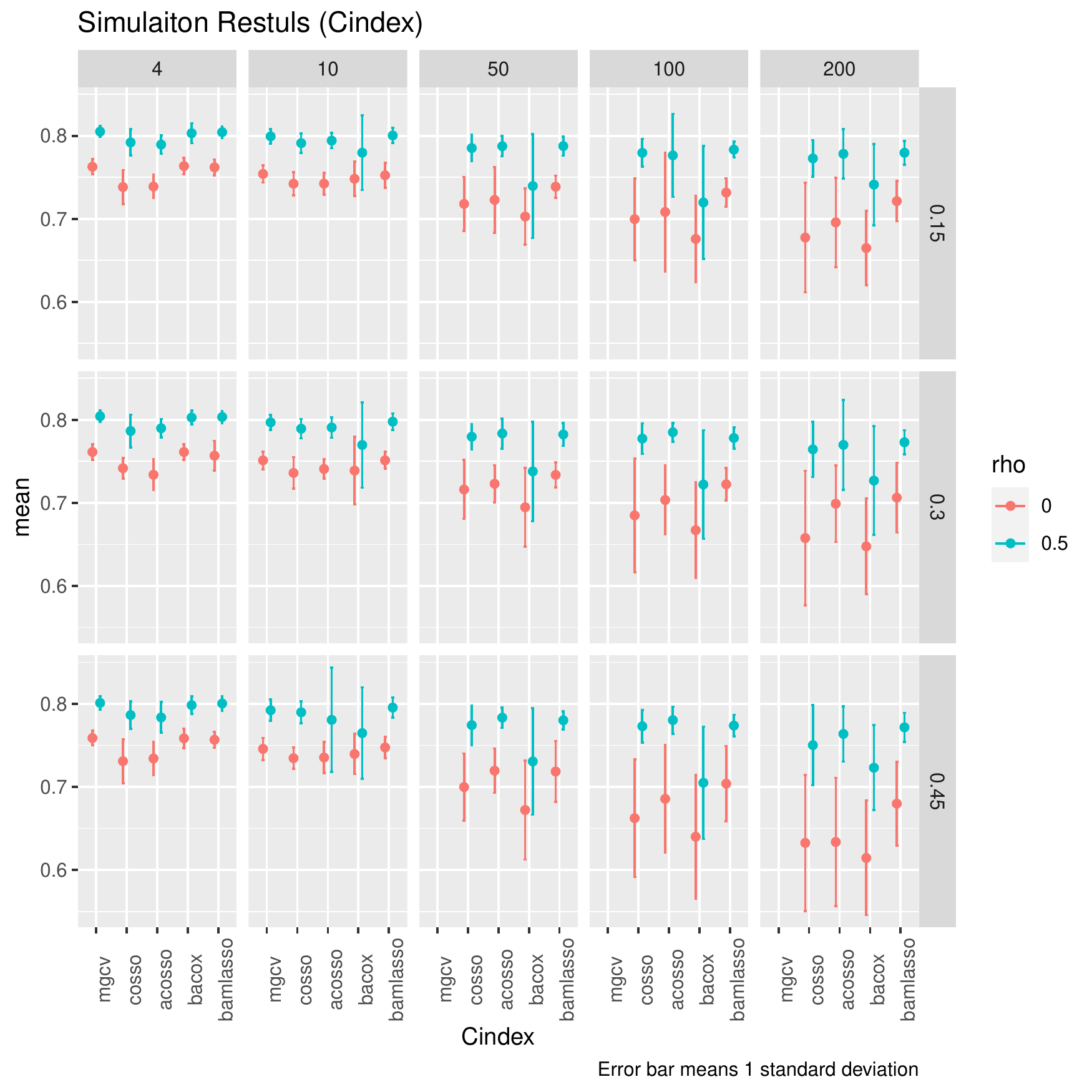}
\caption{C-indices of out-of-sample prediction of 50 iterations of simulation studies. The models of comparison include the proposed Bayesian hierarchical additive model (BHAM), component selection and smoothing operator (COSSO), adaptive COSSO, and gcv. mgcv doesn't provide estimation whe number of parameters exceeds sample size i.e. p = 100, 200.}
\label{fig:sim_cindex}
\end{figure}

Overall, we see consistent performance across different settings of
dimensional, censoring rate and correlation structure (see Figure
\ref{fig:sim_cindex}): the proposed bamlasso model performs as good as,
if not better than previous methods, including mgcv, COSSO, adaptive
COSSO. The improvement is more substantial as p increases, or higher
censoring rate, or when predictors are independent. As expected,
adaptive COSSO performs slightly better than COSSO.

\section{Metabolites Data Analysis}

In this section, we apply the proposed model BHAM to analyze a
previously published metabolically dataset to study plasma metabolomic
profile on the all-cause mortality among patients undergoing cardiac
catheterization. The data set is publicly available via Dryad. The
dataset contains data collected from two cohorts and there exists a
large number of non-overlapping among metabolomic profile measured in
the two cohort. Hence, we focus our analysis on the cohort that have the
larger sample size (N=454). To achieve practically meaningful analysis,
we select the top 200 features with largest variance from the initial
6796 features to conduct our analysis. We use 5-knot cubic splines as
the additive function in the proposed Cox additive model. Optimal tuning
parameters are chosen via 10-fold CV. Out-of-bag samples are used to
evaluate prediction performance, where we primary focus on deviance and
concordance statistics. To visualize the risk prediction, we also create
two groups based on the out-of-bad risk prediction, thresholding at the
median risk and labeled low-risk group and high-risk group. The
Kaplan-Meier plot for the two groups is presented in Figure
\label{ECB_bcam_KM}.

\section{Discussion}

In this article, we introduce the two-part spike-and-slab LASSO prior to
Cox additive models for high-dimensional survival data analysis.
Specifically, the proposed spike-and-slab LASSO prior adopts the sparse
penalty to select effective predictors and the smooth penalty to
estimate the degree of smoothness of the underlying nonlinear function.
This mechanism addresses the previous criticism that high-dimensional
additive models without smooth penalty tends to oversmooth the estimate
of the underlying linear function. Meanwhile, the two-part prior
encourages a bi-level selection of each predictor, selection of
effective predictors and distinguishing linear and nonlinear effects
among predictors. In addition to the Bayesian hierarchical Cox additive
model proposal, we develop a optimization-based algorithm to provide a
scalable solution to fit the model. This substantially reduces
computational demands in comparison to posterior approximation
algorithms. Lastly, we implement the proposed model and fitting
algorithm in a freely available R package \texttt{BHAM} via
\url{https://github.com/boyiguo1/BHAM}, along with many other utility
functions to support model specification, tuning, and diagnostic.

Similar to many machine learning algorithms that provides flexible
modeling of complex signal, the proposed model requires no assumption
checking prior to the model fitting. This is because the model can
automatically choose from the linear or nonlinear forms for a predictor.
Such decisions are data-driven, as the underlying penalty is locally
adaptive, depending on the data. Nevertheless, for the purpose of
variable selection, we still encourage the authors relies on domain
knowledge and consult multiple models. The proposed model can make false
positive selection of linear components. For example, when the
underlying model include no linear component, e.g.~\(x^2\), the proposed
model still tends to include a linear effect in the model. This
phenomenon is because of the effect hierarchy we impose on the prior
structure. In other words, we encourage the model to include the linear
component when the nonlinear effect is detected. An alternative
solution, is to consider the inclusion of the linear and nonlinear
effect independent. Nevertheless, this could leads to lower power to
detect complex functions, i.e.~including both linear and nonlinear
component. Similar trade-offs of different prior construction was
previously seen in \cite{guo2022}.

In this article, we only shows the basic survival model without
mentioning some other analytic problems likely to encounter in survival
analysis, for example delayed entry, stratified Cox models. These
analytic problems can be easily addressed and implemented under the
current model framework. Another important extension of the proposed
model is to model time varying effect. Such extension requires minimum
modification: one can simply construct the additive function matrix for
time, and multiply with the covariates. We see similar strategy was
applied in \cite{wang2007}. In our future work, we attempt to expand the
univariate additive function to multi-dimensional surface, for example
via tensor product spline \cite{Wood2017}. Hence, the interaction effect
between covariates can be modeled. We also aim to extend our framework
to model functional covariates following \cite{cui2021}.

In conclusion, we proposed a new Bayesian hierarchical model to address
Cox additive model in high-dimensional settings, and provide a software
implementation of the proposed model. The model fitting algorithm is
easily scalable and hence can benefit signal detection and risk
prediction in clinical and molecular data analysis. \clearpage
\input{Tabs/sim_cnr_prop.tex} \clearpage

\clearpage

\begin{figure}[h] 
\includegraphics{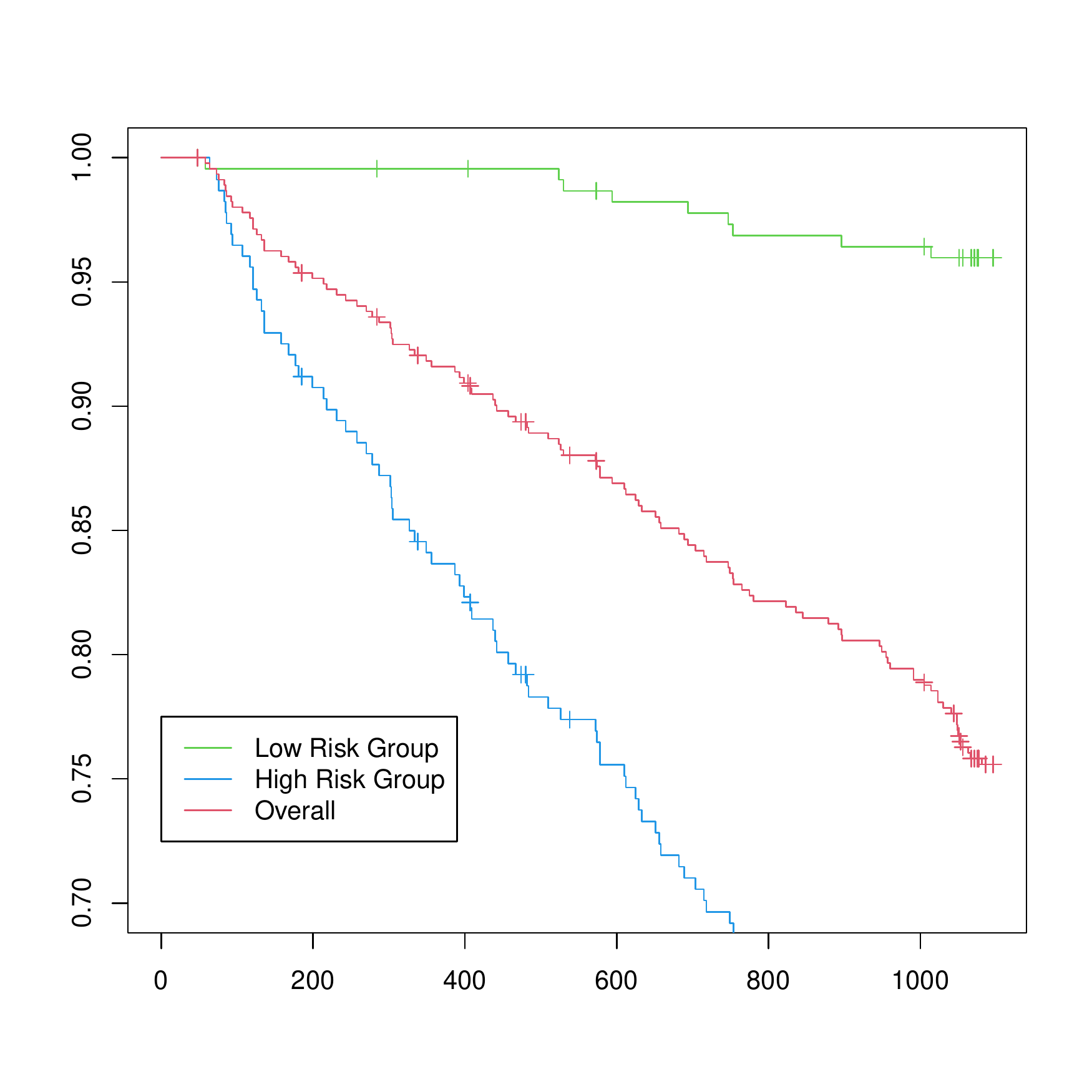}
\caption{The Kaplan-Meier estimates to show the predictive performacne of the proposed model, including the estimates for the predicted high risk group (blue), predicted low risk group (green), and overall group (red)}
\label{fig:ECB_bcam_KM}
\end{figure}

\clearpage

\bibliographystyle{unsrt}
\bibliography{references.bib}

\end{document}

%% file: Tabs/sim_cnr_prop.tex
\begin{table}[ht]
\centering
\begin{tabular}{rrll}
  \hline
sim\_prmt.p & sim\_prmt.pi\_cns & 0 & 0.5 \\ 
  \hline
  4 & 0.15 & 0.155 (0.021) & 0.152 (0.025) \\ 
    4 & 0.30 & 0.304 (0.024) & 0.301 (0.029) \\ 
    4 & 0.45 & 0.465 (0.033) & 0.460 (0.033) \\ 
   10 & 0.15 & 0.155 (0.017) & 0.151 (0.025) \\ 
   10 & 0.30 & 0.310 (0.027) & 0.303 (0.029) \\ 
   10 & 0.45 & 0.463 (0.034) & 0.463 (0.032) \\ 
   50 & 0.15 & 0.151 (0.022) & 0.149 (0.023) \\ 
   50 & 0.30 & 0.303 (0.024) & 0.299 (0.030) \\ 
   50 & 0.45 & 0.458 (0.033) & 0.460 (0.033) \\ 
  100 & 0.15 & 0.146 (0.025) & 0.147 (0.023) \\ 
  100 & 0.30 & 0.306 (0.031) & 0.299 (0.030) \\ 
  100 & 0.45 & 0.462 (0.036) & 0.453 (0.030) \\ 
  200 & 0.15 & 0.149 (0.023) & 0.148 (0.029) \\ 
  200 & 0.30 & 0.302 (0.039) & 0.298 (0.036) \\ 
  200 & 0.45 & 0.464 (0.034) & 0.454 (0.040) \\ 
   \hline
\end{tabular}
\caption{The average and standard deviation of censoring proportion of the simulated data over 50 iterations.} 
\label{tab:sim_cnr_prop}
\end{table}